\documentclass[12pt,preprint]{aastex}

\newcommand{\mdot}{M$_{\odot}$ yr$^{-1}$}
\newcommand{\ldot}{L$_{\odot}$}
\newcommand{\um}{$\mu$m~}

\def\kmsMpc{\ifmmode {\rm\,km\,s^{-1}\,Mpc^{-1}}\else
    ${\rm\,km\,s^{-1}\,Mpc^{-1}}$\fi}

\shorttitle{Most Luminous Starbursts}
\shortauthors{Weedman and Houck}

\begin{document}

\title{The Most Luminous Starbursts in the Universe}

\author{Daniel W. Weedman\altaffilmark{1} and James R. Houck\altaffilmark{1}}

\altaffiltext{1}{Astronomy Department, Cornell University, Ithaca, NY 14853; dweedman@isc.astro.cornell.edu}

\begin{abstract}
A summary of starburst luminosities based on PAH features is given for 243 starburst galaxies with 0 $<$ z $<$ 2.5, observed with the $Spitzer$ Infrared Spectrograph. Luminosity $\nu$L$_{\nu}$(7.7$\mu$m) for the peak luminosity of the 7.7$\mu$m PAH emission feature is found to scale as log[$\nu$L$_{\nu}$ (7.7$\mu$m)] = 44.63($\pm$0.09) + 2.48($\pm$0.28) log(1+z) for the most luminous starbursts observed.  Empirical calibrations of $\nu$L$_{\nu}$(7.7$\mu$m) are used to determine bolometric luminosity $L_{ir}$ and the star formation rate (SFR) for these starbursts. The most luminous starbursts found in this sample have log $L_{ir}$ = 45.4($\pm$0.3) + 2.5($\pm$0.3) log(1+z), in ergs s$^{-1}$, and the maximum star formation rates for starbursts in units of \mdot are log(SFR) = 2.1($\pm$0.3) + 2.5($\pm$0.3) log(1+z), up to z = 2.5.  The exponent for pure luminosity evolution agrees with optical and radio studies of starbursts but is flatter than previous results based in infrared source counts. The maximum star formation rates are similar to the maxima determined for submillimeter galaxies; the most luminous individual starburst included within the sample has log $L_{ir}$ = 46.9, which gives a SFR = 3.4 x 10$^{3}$ \mdot.

\end{abstract}



\section{Introduction}

Understanding the evolution of star formation in the universe is a fundamental objective of observational cosmology.  Various efforts using data from ultraviolet through radio wavelengths have shown that the star formation rate (SFR) per unit volume of the universe increases rapidly with redshift \citep[e.g. ][]{mad98,haa00,cal07,lef05,tak05,man07,mar08}. It is not yet well established, however, why this evolution occurs, at what redshift is the maximum, and whether the evolution in the star formation rate is primarily luminosity evolution (more star formation per galaxy at high redshift) or density evolution (more star-forming galaxies at high redshift). 

Various observations with the Spitzer Space Telescope ($Spitzer$) are providing new insights into this problem, because the signatures of luminous, rapid star formation (starbursts) are easily seen in the spectra obtained with the The Infrared Spectrograph on $Spitzer$ (IRS; Houck et al. 2004).  These features are the strong emission from Polycyclic Aromatic Hydrocarbons (PAH) which dominate the spectra of starbursts in the mid-infrared spectral region from $\sim$ 5\,\um to $\sim$ 20\,\um \citep[e.g. ][]{gen98,rig00,pee04,for04}. These features are remarkably similar in starbursts ranging from low-luminosity, nearby galaxies \citep{bra06} to ultraluminous galaxies at z $\sim$ 2 \citep{pop08,far08}.   

Because extinction corrections are much less significant in the mid-infrared compared to optical or ultraviolet, the PAH luminosities are more confident indicators of intrinsic starburst luminosity than optical or ultraviolet emission.  In addition, the broad wavelength range over which the same features can be consistently observed with the IRS makes it possible to trace PAH luminosity for 0 $<$ z $\la$ 3 using the strong PAH features at 6.2\,\um
and 7.7\,\um.  Observing these features in a wide variety of sources provides a unique tracer of star formation over most of the history of the universe.

In the present paper, we assemble a sample of 243 starbursts selected from a wide variety of IRS observing programs and tabulate their luminosities using a consistent measure of PAH luminosity.  The results trace the most luminous starbursts in the universe at redshifts 0 $<$ z $<$ 2.5 and determine the form of luminosity evolution for starbursts within these redshifts.

\section{Sample Definition and Data Analysis}

\subsection {Sample Selection}

Hundreds of IRS spectra of sources are now available within the $Spitzer$ archive.  Systematic collections of spectra range from studies of nearby, bright ULIRGs, with f$_{\nu}$(24\,\um) $>$ 5\,Jy, to faint sources chosen from various $Spitzer$ surveys with the Multiband Imaging Photometer for $Spitzer$ (MIPS; Rieke et al. 2004), having f$_{\nu}$(24\,\um) $\la$ 1\,mJy.  In this paper, we assemble data from 14 different $Spitzer$ programs to summarize a wide variety of sources having strong PAH spectral features, providing a total of 243 sources.  

While this summary is not a complete review of all IRS spectra with PAH features, it provides a comprehensive data set utilizing a variety of sources chosen for $Spitzer$ observations because of many different criteria.  Although we utilize PAH features as the starburst indicator, the selection criteria for the samples did not incorporate selection effects which arise from the strength or nature of the PAH features.  None of these sources were selected for initial observation based on criteria designed to find sources with strong PAH features.  We have selected from these various observing programs the sources which were subsequently found to have strong PAH features.  The resulting sample provides a diverse set of starbursts which is representative of the full variety of infrared sources which contain starbursts. 

It is especially important to emphasize that this is not a flux limited sample.  Sources were chosen from surveys to many different flux limits and range in f$_{\nu}$(24\,\um) from 8000 mJy (Arp 220) to 0.5 mJy (sources selected from submillimeter surveys).  In samples for which all objects are selected to the same limiting flux, the sources of highest luminosity have the highest redshifts.  In our analysis, we show luminosity trends with redshift, but these do not arise because of comparing objects with similar fluxes but different redshifts. 

The various observing programs by $Spitzer$ used to assemble this collection of IRS spectra of starbursts are now summarized.  The numbers of these observing programs are the same as the reference codes in Table 1, which lists the sample.  The sources used are: 
 
1. 21 bright galaxies classified from optical spectroscopy as pure starbursts and selected for IRS observations because of the initial optical classification as starburst \citep{bra06}.
 
2. 17 starburst galaxies with the strongest PAH features taken from the complete sample of 50 sources in Bootes with f$_{\nu}$(24\,\um) $>$ 10\,mJy in \citet{hou07}.  Sources for this sample were selected based only on the f$_{\nu}$(24\,\um) limit; remaining sources in the 10\,mJy sample have weak or absent PAH features, indicating a substantial AGN contribution to the luminosity. 

3. 19 starburst galaxies in Bootes with the strongest PAH features measured using our own new spectral extractions of $Spitzer$ archival spectra for sources from programs 20128 (G. Lagache, P.I.) and 20113 (H. Dole, P.I.).  These sources were initially selected based on having optical redshifts of 0.2 $\la$ z $\la$ 0.4. 

4. 4 starburst galaxies with the strongest PAH features measured using our own new spectral extractions of a complete sample of 25 sources in the $Spitzer$ First Look Survey \citep{fad06} having f$_{\nu}$(24\,\um) $>$ 10\,mJy (program 40038; Weedman and Houck, in preparation).  Remaining sources in the FLS 10\,mJy sample have weak or absent PAH features.

5. 12 starburst galaxies in the FLS with the strongest PAH features measured using our own new spectral extractions of $Spitzer$ archival spectra for sources from program 20128 (G. Lagache, P.I.).  These sources were initially selected based on optical redshifts of 0.2 $\la$ z $\la$ 0.4.
 
6. 7 starbursts with strong PAH features from the 70\,\um Bootes sample with IRS observations in \citet{brn08}.  Sources were initially selected as a flux limited complete sample of 11 sources with f$_{\nu}$(70\,\um) $>$ 30\,mJy, f$_{\nu}$(24\,\um) $>$ 1\,mJy, and $R$ $>$ 20. 

7. 9 optically faint 24$\mu$m sources from the $Spitzer$ Wide-Area Infrared Extragalactic Survey (SWIRE; Lonsdale 2004) with strong PAH features in IRS spectra \citep{wee06a}. Sources in the Lockman Field were selected as the brightest 24\,\um sources (f$_{\nu}$(24\,\um) $\ga$ 1\,mJy) having estimated photometric redshifts of z $\sim$ 2 as determined from photometry with the $Spitzer$ Infrared Array Camera (IRAC; Fazio et al. 2004).

8. 30 optically faint 24$\mu$m sources from SWIRE fields in \citet{far08}, selected as having f$_{\nu}$(24\,\um) $\la$ 1\,mJy and estimated photometric redshifts of z $\sim$ 2. 

9. 3 optically faint sources with strong PAH features in new extractions of sources with f$_{\nu}$(24\,\um) $\ga$ 1\,mJy in \citet{hou05} or \citet{wee06b}.

10. 15 optically faint 24$\mu$m sources in the FLS having f$_{\nu}$(24\,\um) $\ga$ 1\,mJy with strong PAH features from IRS spectra in \citet{yan07}.  Sources were selected based on IRAC colors. 

11. 12 submillimeter-selected sources with f$_{\nu}$(24\,\um) $\la$ 1\,mJy having strong PAH features in IRS spectra in \citet{pop08}.

12. 36 ultraluminous infrared galaxies (ULIRGs) showing PAH features with IRS spectra in \citet{far07}, chosen from the sample of sources originally discovered by the Infrared Astronomical Satellite (IRAS).

13. 48 IRAS ULIRGS showing PAH features with IRS spectra from \citet{ima07}.

14. 10 IRAS Faint Source Catalog (FSC) sources showing PAH features with IRS spectra in \citet{sar08}; sources were selected as being the optically faintest sources in the FSC.

These sources are listed in Table 1.

\subsection {PAH luminosities for Pure Starbursts}

The characteristic starburst spectrum within the rest wavelengths containing the strongest PAH features observed with the IRS is illustrated by the prototype starburst NGC 7714, shown in Figure 1. All of the strong features arise from PAH emission.  Our objective is to compare luminosities among starbursts over a wide range of redshift using these features.  We desire a straightforward parameter that can be measured with confidence even in sources of poor S/N.  The parameter we choose is the peak luminosity $\nu$L$_{\nu}$ (7.7$\mu$m), where L$_{\nu}$ (7.7$\mu$m) is determined from the flux density at the peak of the 7.7$\mu$m feature.  

The 7.7$\mu$m peak is chosen because this is the strongest PAH feature that is visible in sources at the highest redshifts (z $\sim$ 2.5) for which starburst features   
have been measured with the IRS.  Although the 11.3$\mu$m feature can be comparably strong, it is located at the low-sensitivity edge of IRS spectra at high redshifts.  The dominance of the 7.7$\mu$m feature in spectra of faint, high redshift sources is illustrated in average spectra for such sources in \citet{wee06a}, \citet{pop08}, and \citet{far08}. 

The mixing of the 7.7$\mu$m feature with the adjacent 8.6$\mu$m feature and the difficulty of defining the underlying continuum, especially for sources with silicate absorption centered at 9.7$\mu$m, makes the measurement of total flux in the 7.7$\mu$m feature very uncertain.  The difficulties of measuring total fluxes of the PAH features are discussed, for example, by \citet{pop08} who emphasize that different measurement tools give different answers, depending primarily on the breadth and fitting techniques assumed for the wings of PAH features.  For this reason, we measure only the peak flux density of the feature. 

Uncertainties in the measurement f$_{\nu}$(7.7$\mu$m) for brighter sources, f$_{\nu}$(7.7$\mu$m) $\ga$ 5 mJy, are dominated by uncertainties in the $Spitzer$ flux calibration. This uncertainty is typically $\pm$ 5\%, as estimated by comparing IRS fluxes measured at 24$\mu$m with MIPS fluxes for unresolved sources with MIPS photometry \citep[e.g. ][]{hou07}. For fainter sources, f$_{\nu}$(7.7$\mu$m) $\sim$ 1 mJy, uncertainties are dominated by the low S/N of the spectral peak at 7.7$\mu$m and by resulting uncertainty in fitting the value of the peak.  We have estimated this uncertainty by using different methods of fitting the peak (smooth profile compared to average of the few values near the peak) and conclude that typical uncertainites in f$_{\nu}$(7.7$\mu$m) for the faint sources are $\pm$ 10\%.  

The measures of f$_{\nu}$(7.7$\mu$m) in Table 1 are taken from published values if available; otherwise they are measured from published spectra.  For sources with new extractions, the extractions are derived from v15 of the IRS Basic Calibrated Data and extracted using the SMART analysis package \citep{hig04} as described in \citet{hou07}.  

\citet{hou07} showed that the empirical calibration of $\nu$L$_{\nu}$ (7.7$\mu$m) to SFR using the pure starburst sample of \citet{bra06} leads to a measure of the integrated SFR in the local universe that is consistent with that measured for starburst galaxies using independent techniques.  \citet{hou07} conclude, therefore, that this parameter can be used with confidence to determine the SFR within luminous starburst galaxies, although $\nu$L$_{\nu}$ (7.7$\mu$m) overestimates the starburst luminosity if there is an AGN contribution to the underlying continuum.

The luminosities $\nu$L$_{\nu}$ (7.7$\mu$m) for the sample of 243 galaxies are given in Table 1.  Except for ULIRGs chosen from bright IRAS sources, the galaxies are included only if the references classify the sources as starbursts, based on the equivalent widths of the PAH features or on fitting the spectrum with a pure starburst template.  For these "pure starbursts", we have measured f$_{\nu}$ (7.7$\mu$m) from published spectra or from new extractions of unpublished archival data, and these measured values are included in Table 1.  For ULIRGs, discussed further below in section 2.3, the $\nu$L$_{\nu}$ (7.7$\mu$m) are scaled to published luminosities for other PAH features because there may be a significant contribution to the $\nu$L$_{\nu}$ (7.7$\mu$m) from an underlying AGN continuum. 

Luminosities in Table 1 range over a factor of $>$ 10$^{4}$, from log[$\nu$L$_{\nu}$ (7.7$\mu$m)] = 41.76 for no. 150, a blue compact dwarf from the Bootes 10 mJy survey in \citet{hou07}, to log[$\nu$L$_{\nu}$ (7.7$\mu$m)] = 46.10 for no. 197, a faint source from the FLS survey in \citet{yan07}. All luminosities are shown compared to redshift in Figure 3. 

\subsection {PAH Luminosities for Composite ULIRGs}

The only sources included in this summary which sometime have spectroscopic indicators of an AGN contribution are the IRAS ULIRGs, which often show the deep silicate absorption characteristic of AGN while also showing PAH features \citep[e.g. ][]{spo07} .  For such composite sources, simply taking $\nu$L$_{\nu}$ (7.7$\mu$m) would not be an appropriate measure of starburst luminosity because $\nu$L$_{\nu}$ (7.7$\mu$m) can be enhanced by an underlying AGN continuum.
 
These ULIRGs are sufficiently bright to have S/N adequate for measurement of total fluxes in the weaker PAH features at 6.2$\mu$m and/or 11.3$\mu$m.  These two PAH features are sufficiently isolated in the spectrum that they can be measured as single Gaussians on top of an underlying continuum, so that a total flux can be determined much more precisely than for the 7.7$\mu$m feature.  The blending of PAH features makes especially difficult the definition of total flux in the 7.7$\mu$m feature, as discussed, for example, by \citet{bra06} and \citet{pop08}.  Furthermore, sources with deep 9.7$\mu$m silicate absorption have a continuum peak near 8$\mu$m which further complicates isolation of the 7.7$\mu$m feature.  This peak is seen in the spectrum of Markarian 231, shown in Figure 1. 

The effects of starburst and AGN composite spectra are illustrated in Figure 1.  Rest-frame spectra are shown only between 5$\mu$m and 14$\mu$m because this is the spectral region with diagnostics that differentiate starbursts and AGN.  Also, these rest-frame wavelengths are all that can be observed with the IRS in sources with z $\ga$ 1.5, which are the high redshifts of particular importance to our analysis. The AGN Markarian 231 has an absorption spectrum and luminosity that is typical of absorbed ULIRGs \citep{sar08}, so this ULIRG provides an excellent representative example for an AGN contribution to the mid-infrared spectrum.  For a starburst comparison to Markarian 231, we use the IRS spectrum from \citet{bra06} of the prototype starburst NGC 7714.  

Figure 1 illustrates composite spectra that arise from different mixtures of the Markarian 231 and NGC 7714 spectra, from pure starburst at the top through starburst contributions of 75\%, 50\%, and 25\% to pure AGN at the bottom.  Spectra are normalized to the peak which is between 7.7$\mu$m and 7.9$\mu$m, depending on the composite.  The decreasing equivalent width of the PAH features is evident as the starburst contribution lessens.  This is shown quantitatively in Figure 2, which tracks the equivalent width (EW) in the rest frame of the 6.2$\mu$m feature as a function of the starburst contribution.  This EW is measured by fitting a single Gaussian profile to the 6.2$\mu$m feature and using the underlying "continuum" defined between 5.5$\mu$m and 6.9$\mu$m.  For pure starbursts, much of this "continuum" may actually be extended wings of the 7.7$\mu$m PAH complex, but measuring the 6.2$\mu$m feature in this way provides a consistent measurement of its strength. 

The "pure starbursts" in our sample generally have rest frame equivalent widths for the 6.2$\mu$m feature greater than 0.5$\mu$m, indicating that the starburst contribution exceeds 90\% of the mid-infrared luminosity.  However, equivalent widths for the ULIRGs are generally less than this value, implying an AGN contribution to the underlying continuum.  For the ULIRGs, a simple measure of $\nu$L$_{\nu}$ (7.7$\mu$m) could overestimate the starburst luminosity.  For this reason, we determine a $\nu$L$_{\nu}$ (7.7$\mu$m) for the starburst component of ULIRGs by using the 
published luminosities of the total PAH features L(6.2$\mu$m) \citep{ima07,sar08} or L(6.2$\mu$m+11.3$\mu$m) \citep{far07}.  Such luminosities are measured only for the PAH emission feature and do not include any continuum contribution from the AGN.  

Luminosities of these features are related to the luminosity of our PAH parameter $\nu$L$_{\nu}$(7.7\,\um) using empirical transformations which have been determined by measuring $\nu$f$_{\nu}$ (7.7$\mu$m), f(6.2$\mu$m), and f(11.3$\mu$m) for the brightest 69 pure starbursts in Table 1 (Sargsyan et al., in preparation).  These transformations give that $\nu$L$_{\nu}$(7.7\,\um) = 50($\pm$15) L(6.2\,\um) and $\nu$L$_{\nu}$(7.7\,\um) = 28($\pm$6) L(6.2$\mu$m+11.3$\mu$m).  Using these transformations, the equivalent $\nu$L$_{\nu}$ (7.7$\mu$m) for ULIRGs are listed in Table 1.

 
\section{Discussion}

\subsection{Redshift Distribution of PAH Luminosities} 

PAH luminosities $\nu$L$_{\nu}$(7.7\,\um) of the sources in Table 1 are plotted in Figure 3.  The distribution of luminosities with redshift shows a well defined upper envelope for starburst luminosity.  Because the redshift intervals are not evenly sampled, we have quantitatively fit this envelope to avoid the sampling differences by choosing the most luminous sources in each interval of 0.02 in log(1+z).  There are 27 such intervals in the redshift range plotted, and all intervals have sources within the interval except for 4 intervals (0.3 to 0.32, 0.32 to 0.34, 0.36 to 0.38, and 0.52 to 0.54).  The linear (first order) least squares fit to these most luminous sources is given by \begin{equation}$$log[$\nu$L$_{\nu}$ (7.7$\mu$m)] = 44.63($\pm$0.09) + 2.48($\pm$0.28) log(1+z)$$\end{equation}
The solid line illustrated in Figure 3 illustrates this fit for maximum starburst luminosity for 0 $<$ z $<$ 2.55.
  
The luminosities $\nu$L$_{\nu}$(7.7\,\um) in Table 1 and Figure 3 contain no extinction corrections.  For pure starbursts, these corrections would be small.  The sample in \citet{bra06} has an average optical depth at the 9.7\,\um silicate absorption feature of 0.24.  Using the extinction curve of \citet{dra01} and assuming this silicate absorption also applies to the PAH features, the extinction at the 7.7\,\um feature would be only 0.05 mag.  Sources in our sample which would be most vulnerable to uncertain extinction corrections are the IRAS ULIRGs, because their spectra often shown clear evidence of substantial silicate absorption by silicate dust \citep[e.g. ][]{spo07, ima07, des07}. The starburst luminosities of IRAS ULIRGs are shown with different symbols (triangles) in Figure 3 and are discussed more below.  

While extinction corrections might affect the results for $\nu$L$_{\nu}$(7.7\,\um) for some sources in Figure 3, this would not affect our  primary conclusions.  This is because our most important use of the PAH luminosities is in section 3.3, where they are transformed to bolometric luminosities ($L_{ir}$) and star formation rates (SFR).  This transformation is empirical, so extinction corrections are not relevant for determining $L_{ir}$ and SFR as long as the systematic corrections are the same in the sources used to determine the empirical transformations as in the sources to which the transformations are applied.

\subsection{Comparisons with IRAS ULIRGs}

Of the 243 sources in Table 1, 83 (34\%) are included because they had been selected as ULIRGs.  To allow identification of these IRAS ULIRGs, Figures 3 and 4 use different symbols (triangles) to show the luminosities of the starburst component in the IRAS ULIRG samples compared to the pure starbursts from other samples. (The IRAS FSC sources from \citet{sar08} are also shown with ULIRG symbols although they were not included because of initial selection as ULIRGs.)

For z $<$ 0.2 where most ULIRGs are found, the starbursts of maximum luminosity are primarily in ULIRGs, although the most luminous ULIRG starbursts are no more luminous than the most luminous pure starbursts. For sources over all redshifts shown in Figure 4, the median luminosities of ULIRG starbursts and pure starbursts are very similar.  These results indicate that the starbursts in absorbed ULIRGs are not systematically different from starbursts in other sources.     

Many previous studies of ULIRGs show that these sources often contain some luminosity from an AGN \citep{far07,ima07,des07} and that many have buried luminosity sources which are extremely obscured  \citep[e.g. ][]{lev07}.  With sufficient obscuration, no direct spectroscopic indicators can emerge from the obscuring dust to determine if the obscured source shows the characteristic PAH features of starbursts or the high ionization lines of AGN. \citet{ima07} argue from radiative transfer considerations that the buried sources are most probably AGN which heat their immediate surroundings to a high dust temperature, and this hot continuum is progressively absorbed by cooler surrounding dust.  

Such a scenario leads to a consistent description of those ULIRGs showing both strong silicate absorption and PAH features.  This description is that the buried AGN is the obscured source and that the PAH features arise from much less obscured starbursts occuring within the clouds surrounding the AGN. In this case, the extinction to the starburst would be much less than the extinction derived from the optical depth of the silicate feature.

There is qualitative evidence supporting this description.  This evidence is the similarity in the ratio of the 11.3\,\um to 6.2\,\um PAH fluxes in composite spectra of heavily absorbed ULIRGs compared to this ratio in pure starbursts.  This result can be seen in Figure 1 of \citet{des07} which compares the average starburst spectrum from \citet{bra06} to the average spectrum of heavily absorbed ULIRGs.  The 11.3\,\um to 6.2\,\um ratio is approximately 1.0 in both cases, even though the 11.3\,\um feature is within the 9.7\,\um silicate absorption so should be suppressed if it is obscured by the same silicate absorption that obscures the buried source.  

The average absorbed ULIRG shown in \citet{des07} has a 9.7\,\um silicate optical depth of 1.5.  For this absorption,  the 11.3\,\um feature is extincted by 1.1 mag compared to 0.48 mag of extinction for the 6.2\,\um feature, using the extinction curve of \citet{dra01}.  This differential extinction would imply that the observed 11.3\,\um to 6.2\,\um ratio would be 0.56 for an intrinsic ratio of 1.0.  That such a lower ratio is not observed in the absorbed ULIRGs is evidence that their starbursts are not extincted by the same dust that produces the silicate absorption.

This reasoning justifies treating ULIRG starbursts the same as other starbursts regarding extinction corrections.  Therefore, we do not apply extinction corrections to the ULIRGs and will apply the same transformation between PAH luminosities and $L_{ir}$ to the ULIRG starbursts as to the other starbursts.

\subsection{Bolometric Luminosities and Star Formation Rates}

The plotted values and fitted envelope in Figure 3 derive strictly from observed data, with no assumptions regarding templates or spectral shapes for starburst galaxies.  The only assumption is that the values of $\nu$L$_{\nu}$ (7.7$\mu$m) arise purely from a starburst, with no contribution from an AGN.  As discussed above, the sample was chosen in order to use only pure starbursts with no evidence that the PAH complex is diluted by an AGN, or to use sources with published luminosities of individual PAH features for the starburst component when there is evidence of an AGN.

The $\nu$L$_{\nu}$ (7.7$\mu$m) in Figure 3 can be transformed to bolometric luminosities ($L_{ir}$) and star formation rates (SFR) using empirically determined conversions.  Such conversions and their relevant uncertainties are subject to further refinement, but this would not affect the data shown in Figure 3.  For further discussion, the conversions which are adopted are those from \citet{hou07} for pure starbursts. These conversions are: \begin{equation}$$log $L_{ir}$ = log[$\nu$L$_{\nu}$ (7.7$\mu$m)] + 0.78($\pm$0.2)$$\end{equation} for $L_{ir}$  in ergs s$^{-1}$, and \begin{equation}$$log[SFR] = log[$\nu$L$_{\nu}$ (7.7$\mu$m)] - 42.57($\pm$0.2)$$\end{equation}for SFR in \mdot.

Equation (3) is derived using the relation from \citet{ken98} between $L_{ir}$ and SFR. \citet{hou07} showed that the conversions in equations (2) and (3), which are derived empirically from the sample of 22 starburst galaxies in \citet{bra06}, lead to results for the local SFR density that are consistent with independent results from all sky samples of IRAS which determine $L_{ir}$ using measured continuum luminosities from IRAS. 

The transformations in equations (2) and (3) have no dependence on luminosity because there is no evidence of such dependence in the starburst sample of \citet{bra06}.  \citet{pop08} derive empirical transformations between PAH luminosities and $L_{ir}$ using $L_{ir}$ determined for high redshift starbursts selected from submillimeter detections and subsequently observed spectroscopically with IRS.  These starbursts are typically $\sim$ 100 times more luminous than the local starbursts of \citet{bra06}, but the relation between $L_{ir}$ and PAH luminosities is an extension of the relation from the local Brandl starbursts, with no dependence on luminosity. The agreement of this calibration for high luminosity, high redshift sources with the calibration for the local, lower luminosity starbursts in \citet{bra06} is the strongest evidence we have that the calibration is not luminosity dependent.  This lack of luminosity dependence is also emphasized by \citet{pop08}.

Uncertainties given in equations (2) and (3) are the random uncertainties among the Brandl starburst sample used for the conversion; systematic uncertainties in the adopted transformations can be estimated by comparison with independent determinations of the same parameters. Such comparisons are now discussed. 

>From combined local and submillimeter-derived samples, \citet{pop08} relate $L_{ir}$ and PAH luminosity by log $L_{ir}$ = log[L(6.2$\mu$m)] + 2.7$\pm$0.1.  The transformation  between $\nu$L$_{\nu}$(7.7$\mu$m) and L(6.2$\mu$m) that we discuss in section 2.3 from starbursts in our sample is log[$\nu$L$_{\nu}$ (7.7$\mu$m)] = log[L(6.2$\mu$m)] + 1.70($\pm$0.13).  Using this in conjunction with equation (2), we would have the result that log $L_{ir}$ = log[L(6.2$\mu$m)] + 2.5.  
Within the uncertainties, this is the same as the relation in \citet{pop08}. 

Earlier studies have also related broad-band photometric luminosities to  bolometric luminosities.  This has been done for various monochromatic wavelengths, and estimates are available at rest-frame 8$\mu$m \citep{cap07,bav07}.  Applying a transformation from L(8$\mu$m) to $L_{ir}$ for large samples of infrared sources with only photometric data requires assumptions about the mix of spectroscopic templates to be used.  The transformation from near-infrared to bolometric luminosities depends on whether a source is an AGN or a starburst and what mix of spectra applies to the ensemble of sources. 
 
\citet{cap07} and \citet{bav07} derive transformations based on observed fluxes at 8$\mu$m, 24$\mu$m, 70$\mu$m, and 160$\mu$m taken with a mix of templates. At 8$\mu$m, the \citet{cap07} result is log $L_{ir}$ =  1.06 log[$\nu$L$_{\nu}$ (8$\mu$m)] + 0.28, for $\nu$L$_{\nu}$ (8$\mu$m) in \ldot.  The \citet{bav07} result is  log $L_{ir}$ =  0.83 log[$\nu$L$_{\nu}$ (8$\mu$m)] + 2.57, also using \ldot.  We can compare these transformations to those we use based on $\nu$L$_{\nu}$(7.7$\mu$m) at the peak of the PAH feature.

Because of slightly different luminosity dependencies among the transformations in \citet{cap07} and \citet{bav07}, we use for comparisons a representative luminosity for the most luminous starbursts of log[$\nu$L$_{\nu}$ (7.7$\mu$m)] = 45.5, in ergs s$^{-1}$, or 11.9 in \ldot.   At this luminosity, our transformation gives log $L_{ir}$ = 46.3 in ergs s$^{-1}$, or log $L_{ir}$ = 12.7 in \ldot. Taking the $\nu$L$_{\nu}$ (8$\mu$m), \citet{cap07} would derive log $L_{ir}$ = 12.9, and \citet{bav07} would derive 12.4.  

These results bracket our result based on the empirical transformation for pure starbursts, so this is independent evidence that within the uncertainties of the various estimates, our transformation is valid.  This consistency also indicates that the \citet{cap07} and \citet{bav07} transformations apply primarily for starbursts rather than for AGN.  The weak dependence on luminosity in the \citet{cap07} and \citet{bav07} results is also evidence that justifies our application of a luminosity-independent transformation.

The transformation from a PAH luminosity to a SFR depends on relating the SFR to the bolometric luminosity, $L_{ir}$, after adopting a relation between PAH luminosity and $L_{ir}$.  Such transformations are most commonly based on the relations in \citet{ken98} which determine the bolometric luminosity from a star-forming galaxy as a function of the total SFR.  The bolometric luminosity can be related to various other measured parameters, such as ultraviolet luminosity from the ionizing stars or emission line luminosities.  Results can be refined for different initial mass functions, starburst ages and estimates of extinction.

For example, the SFR can be determined from the luminosity of the [Ne II] and [Ne III] lines \citep{ho07}.  By empirically comparing luminosities of the Neon lines with those of PAH features, \citet{far07} relate the SFR as given by \citet{ho07} to the luminosities of the 6.2$\mu$m plus 11.3$\mu$m PAH features in a starburst. The result in \citet{far07} is log(SFR) = log[L(6.2$\mu$m + 11.3$\mu$m)] - 40.9.   If we use our transformation discussed in section 2.3, $\nu$L$_{\nu}$(7.7\,\um) = 28($\pm$6) L(6.2$\mu$m+11.3$\mu$m), our equation (3) becomes log(SFR) = log[L(6.2$\mu$m + 11.3$\mu$m)] - 41.1.  To within the uncertainties, this is the same result as derived by \citet{far07} from the Neon lines.  

For now, therefore, we consider the relations in equations (2) and (3) for determining bolometric luminosity and SFR to be as accurate as other methods currently available, although we certainly expect refinements in the future.  Using these transformations along with equation (1) for the change of $\nu$L$_{\nu}$(7.7\,\um) with redshift as shown in Figure 3, we then have that\begin{equation}$$log $L_{ir}$ = 45.4($\pm$0.3) + 2.5($\pm$0.3) log(1+z)$$\end{equation}for $L_{ir}$ in ergs s$^{-1}$, and\begin{equation}$$log(SFR) = 2.1($\pm$0.3) + 2.5($\pm$0.3) log(1+z)$$\end{equation} for SFR in \mdot.

The most luminous starburst in the sample is MIPS 506 \citep{yan07}, no. 192 in Table 1, with log[$\nu$L$_{\nu}$ (7.7$\mu$m)] = 46.10, which gives a SFR = 3.4 x 10$^{3}$ \mdot.

We emphasize again that these results are based strictly on the measurement of PAH luminosity in starburst sources.  It is useful, therefore, to compare with previous results for maximum SFR derived from completely independent selection and measurement criteria.  The most luminous starburst galaxies previously reported are those in the submillimeter galaxy population.  While this population contains both AGN and starbursts, sources with cooler dust are attributed primarily to starbursts.  The upper envelope for log($L_{ir}$) for submm-discovered starbursts is $\sim$ 46.5 at z $\sim$ 2 \citep{cha05}.  This result is very similar to the maximum luminosity of 46.6 which we would find for starbursts at z = 2 from equation (4).   This comparison indicates, therefore, that the use of PAH features for seeking luminous starbursts does indeed find sources that are as luminous in bolometric luminosity as those known from other techniques.

\subsection{Starburst Evolution}


The evolution of star formation in the universe is fundamentally described by determining the change of the total luminosity density of the universe as a function of redshift \citep[e.g. ][]{mad98,lef05,mar08}.  This can be described by various combinations of pure luminosity evolution and pure density evolution for star-forming galaxies.  Pure luminosity evolution means that the shape of the luminosity function for star-forming galaxies does not change; the number of galaxies per comoving volume remains the same and only their luminosities change.  All galaxies at a given space density in the luminosity function simply scale up in luminosity by the same factor with increasing redshift. 

For pure luminosity evolution, the evolution of the entire luminosity function is the same as the evolution of the most luminous starbursts.  With such evolution, results determined above in section 3.1 which describe the scaling of luminosity with redshift would apply to all star-forming galaxies. We cannot conclude that such evolution applies to lower luminosity star formation, however, because our data are not sufficient to determine the shape of the luminosity function at different redshifts. 

The thorough study of \citet{lef05} using photometric redshifts of $Spitzer$-detected sources and template fitting to determine source luminosities indicates a factor of (1+z)$^{3.9}$ for evolution of the total luminosity density to z $\sim$ 1.  \citet{lef05} separate this into a factor of (1+z)$^{3.5}$ for luminosity evolution and (1+z)$^{0.5}$ for density evolution.  This small factor for density evolution is evidence that the number density of star forming galaxies changes little; evolution is primarily in the luminosity per galaxy.  These results for total evolution also agree within the uncertainties with earlier estimates based on infrared \citep{lag04} and submillimeter \citep{cha05} source counts and redshifts. 

The comprehensive summary by \citet{hop04} of evolution parameters derived from optical and radio constraints gives smaller values, with evolution going as (1+z)$^{2.9}$ and being almost entirely luminosity evolution.  Hopkins states that the preferred exponent for luminosity evolution is 2.7 $\pm$0.6. This form of evolution continues to z $\sim$ 2.5 but may flatten beyond that redshift.  Because of the uncertainty in the exponent for the (1+z) factor, the difference in the luminosity evolution exponent between optical/radio results and infrared/submillimeter results may not be significant. 

We can compare the previously determined evolution parameters with those we determine from the most luminous sources.  For our results, the envelope describing maximum starburst luminosity to z = 2.5 shown in Figures 3 and 4 has the maximum luminosity of starbursts evolving as (1+z)$^{2.5}$.  For lower redshifts, to z = 1, another envelope is shown with the \citet{lef05} factor for luminosity evolution of (1+z)$^{3.5}$.  Although the uncertainties in the exponents indicate that our evolution factor and the \citet{lef05} factor are similar within uncertainties, the steeper evolution does not track well the lower redshift starbursts in Figures 3 and 4 and would certainly overestimate luminosities at higher redshifts.  

Our results for the summary of starbursts in Figure 3 indicate, therefore, that luminosity evolution for the most luminous starbursts scales more closely to the lower value of \citet{hop04}. Figure 3 is also a clear demonstration that this luminosity evolution continues at least to z = 2.5. A possible explanation for the differences found for evolution of pure starbursts compared to the evolution derived from modeling total infrared source counts and redshift distributions, such as in \citet{lef05}, is that infrared sources include a significant fraction of AGN, and the evolution of AGN may differ from that of starbursts. 

\section{Summary and Conclusions}

A sample of 243 starburst galaxies with infrared spectra obtained by The Infrared Spectrograph on $Spitzer$ has been assembled with measurements of PAH luminosities to determine the most luminous starbursts discovered.  The sample includes sources from a variety of $Spitzer$ observing programs and covers 0 $<$ z $<$ 2.5 (Figure 3).

Starburst luminosities are derived from the luminosity $\nu$L$_{\nu}$ (7.7$\mu$m) as determined from the peak flux density of the 7.7$\mu$m PAH feature. These luminosities for the most luminous starbursts scale with redshift as log[$\nu$L$_{\nu}$ (7.7$\mu$m)] = 44.63($\pm$0.09) + 2.48($\pm$0.28) log(1+z). This result demonstrates that pure luminosity evolution for starbursts scales approximately with (1+z)$^{2.5}$, at least to z = 2.5.  This is less evolution than determined from previous infrared-derived source counts but agrees with the evolution determined from optical and radio samples of star-forming galaxies.

Transformations of $\nu$L$_{\nu}$ (7.7$\mu$m) to bolometric luminosities $L_{ir}$ and to star formation rates are determined empirically from local starbursts and are shown to be the same as such transformations derived by others using a variety of star formation indicators and a variety of sources.  
Using the conversions that log $L_{ir}$ = log[$\nu$L$_{\nu}$ (7.7$\mu$m)] + 0.78, and that log[SFR] = log[$\nu$L$_{\nu}$ (7.7$\mu$m)] - 42.57, for luminosities in ergs s$^{-1}$ and SFR in \mdot, we find that:

1. Bolometric luminosities of the most luminous starbursts in the universe scale with redshift as log $L_{ir}$ = 45.4($\pm$0.3) + 2.5($\pm$0.3) log(1+z).  

2. The SFR of the most luminous starbursts in the universe scales with redshift as log(SFR) = 2.1($\pm$0.3) + 2.5($\pm$0.3) log(1+z), to z = 2.5. 

The most luminous starbursts in the sample are similar in SFR to the most luminous starbursts previously found in submillimeter surveys; the maximum starburst in the sample has SFR = 3.4 x 10$^{3}$ \mdot.   

We also find that at the redshifts of IRAS ULIRGs, z $<$ 0.2, the most luminous ULIRG starbursts are similar in luminosity to the most luminous pure starbursts.  These results indicate that the starburst component in composite, heavily absorbed ULIRGs having both a starburst and  AGN component is not systematically different from the pure starbursts in other sources.


\acknowledgments
We thank L. Sargsyan for help with data analysis. This work is based primarily on observations made with the
Spitzer Space Telescope, which is operated by the Jet Propulsion
Laboratory, California Institute of Technology, under NASA contract
1407. Support for this work by the IRS GTO team at Cornell University was provided by NASA through Contract
Number 1257184 issued by JPL/Caltech.

\clearpage


\begin{deluxetable}{ccccccc} 
\tablecolumns{7}
\tabletypesize{\footnotesize}

\tablewidth{0pc}
\tablecaption{PAH Luminosities for Starburst Galaxies}
\tablehead{
 \colhead{No.} &\colhead{Name}& \colhead{J2000 coordinates} &\colhead{z} & \colhead{f$_{\nu}$(7.7$\mu$m)\tablenotemark{a}}& \colhead{log $\nu$L$_{\nu}$(7.7\,\um)\tablenotemark{b}}& \colhead{Ref.\tablenotemark{c}} 
}
\startdata

1 & IRAS 00091-0738  &001143.24-072207.3 & 0.118 & \nodata & 43.81 & 13\\ 
2 & IRAS 00188-0856 & 002126.48-083927.1  & 0.128 & \nodata & 44.24 & 13\\
3 & IRAS 00397-1312 & 004215.50-125603.5	 & 0.262 & \nodata & 45.20& 12\\
4 & IRAS 00456-2904  &004806.75-284818.6 & 0.110&\nodata & 44.74 & 13\\
5 & IRAS 00482-2721  &005040.41-270438.3  & 0.129 &\nodata& 43.95 & 13\\
6 & IRAS 01004-2237  &010249.94-222157.3  & 0.118 &\nodata & 44.19 & 13\\
7 & IRAS 01166-0844  &011907.55-082909.4   & 0.118 & \nodata & 43.40 & 13\\
8  & NGC 520 & 012435.07+034732.7	&0.0071	& 1120	&43.83 & 1\\ 
9  & IRAS 01298-0744  &013221.41-0729.08.5  & 0.136 & \nodata & 43.54 & 13\\
10 & NGC 660 & 014302.35+133844.4 & 0.0029 &2020	&43.31& 1\\  
11 & IRAS 01569-2939  &015913.79-292434.5  & 0.141 & \nodata & 44.19 & 13\\
12 & IRAS 02411+0353  &024346.09+040636.9 & 0.144 & \nodata & 44.81 & 13\\
13 & NGC 1097 & 024619.08-301628.0 &0.0040 &1120	&43.86 & 1\\ 
14 & NGC 1222 &030856.74-025718.5& 0.0076 &330 & 43.34& 1\\
15 & IRAS 03158+4227 & 031912.60+423828.0	& 0.134 & \nodata & 44.54 & 12\\
16 & IRAS 03521+0028 & 035442.15+003702.0	& 0.152 & \nodata & 44.56& 12\\
17 & IRAS 03250+1606  & 032749.77+161659.8 &0.129 &\nodata& 44.54 & 13\\
18 & IC 342 & 034648.51+680546.0	&0.0011	&7200  &42.67 & 1\\ 
19 & IRAS 04103-2838   &041219.53-283024.4  &  0.118 &\nodata& 44.62 & 13\\ 
20 & NGC 1614 &043359.85-083444.0	 & 0.0148 & 940 & 44.53&  1\\ 
21 & IRAS 05189-2524 & 052101.41-252145.5	 & 0.043 & \nodata & 44.25& 12\\
22 &  IRAS 06035-7102 & 060253.63-710311.9	 & 0.079 & \nodata & 44.57&  12\\
23 & NGC 2146 &061837.71+782125.3	 & 0.0039 & 4800	 & 44.06&  1\\ 
24 & IRAS 06206-6315 & 062100.80-631723.2	 & 0.092 & \nodata & 44.47 &  12\\
25 &  FSC 07247+6124 &072912.10+611853.5 & 0.137 &\nodata&43.85 & 14\\
26 &  IRAS 07598+6508 & 080430.46+645952.9	 & 0.148 & \nodata & 44.44&  12\\
27 & NGC 2623 &083824.08+254516.9	 & 0.0183 & 340	 & 44.02 & 1\\ 
28 & IRAS 09022-3615 & 090412.00-362701.0	 & 0.060 & \nodata & 44.76&  12\\
29 & IRAS 09039+0503  &090634.05+042126.0   & 0.125  &\nodata& 44.27 & 13\\
30 & IRAS 09116+0334   & 091413.82+032201.3  & 0.146 &\nodata& 44.67 & 13\\
31 &  FSC 09235+5425 &092703.07+541206.6  &0.123  &\nodata&43.95 & 14\\ 
32 &  FSC 09284+0413 & 093101.27+035955.2 &0.146 &\nodata& 44.53 & 14\\
33 & UGC 5101 & 093551.65+612111.3	 & 0.039 & \nodata & 44.63& 12\\
34 & IRAS 09539+0857  &095634.35+084305.6    & 0.129 &\nodata& 44.09 & 13\\
35 & IRAS 10190+1322W  & 102142.56+130653.3   & 0.077 &\nodata& 44.36 & 13\\
36 & IRAS 10190+1322E  &102142.81+130655.0   & 0.077 &\nodata& 44.34 & 13\\
37 &  FSC 10219+2657  &102447.39+264209.0 &0.225  &\nodata&44.61  & 14\\
38 & NGC 3256 &102751.27-435413.8	 & 0.0084 & 2100 & 44.22& 1\\ 
39 &SWIRE& 103205.16+574817.5 & 1.63 & 1.3 & 45.53   & 8\\  
40 &SWIRE& 103707.80+591204.5 & 1.84 & 1.0 & 45.52 &   8\\  
41 &  SWIRE& 103744.46+582950.6	 & 1.88 & 	1.6 & 45.73 &  7\\
42 & IRAS 10378+1108   &104029.17+105317.7   & 0.136 &\nodata& 44.04 & 12\\ 
43 &  SWIRE& 103809.18+583226.2 & 	0.98 & 	1.4 & 45.13 &   7\\
44 &  SWIRE& 103837.03+582214.8	 & 1.68 & 	2.5 & 45.84 &   7\\
45 & NGC 3310 &103845.96+533005.3	 & 0.0047 & 400	 & 43.42&  1\\ 
46 &  SWIRE& 103856.98+585244.1	 & 1.88 & 	1.5 & 45.70 &   7\\
47 &SWIRE& 104011.61+580542.7 & 1.83 & 1.7 & 45.74 &  8\\  
48 &SWIRE& 104012.86+592712.5 & 1.43 & 3.3 & 45.82 &   8\\  
49 & IRAS 10378+1109 & 104029.17+105317.7 & 	0.136 & \nodata & 44.37& 13\\
50 &SWIRE& 104034.35+582314.6 & 1.89 & 1.3 & 45.64 &  8\\  
51 &SWIRE& 104129.26+581712.3 & 1.61 & 1.7 & 45.65 &   8\\  
52 &SWIRE& 104139.78+573723.9 & 1.69 & 1.6 & 45.65 &   8\\  
53 &  SWIRE& 104217.17+575459.2	 & 1.91 & 	1.7 & 45.77 &   7\\
54 &SWIRE& 104232.04+575439.5 & 1.93 & 1.6 & 45.75 &  8\\  
55 &SWIRE& 104343.93+571322.5 & 1.71 & 2.2 & 45.80 &   8\\  
56 &SWIRE& 104349.42+575438.7 & 1.77 & 1.1 & 45.52 &   8\\  
57 &SWIRE& 104402.56+593204.7 & 1.85 & 1.1 & 45.57 &   8\\  
58 &SWIRE& 104427.54+593811.8 & 1.83 & 1.2 & 45.59 &   8\\ 
59 &SWIRE& 104436.56+593252.4 & 1.49 & 1.8 & 45.60 &   8\\  
60 &SWIRE& 104514.39+575708.9 & 1.79 & 1.5 & 45.68 &   8\\  
61 &SWIRE& 104614.90+594134.3 & 2.11 & 1.3 & 45.55   & 8\\ 
62 &  SWIRE& 104620.38+593305.1	 & 1.84 & 	2.0 & 45.81 &   7\\
63 &SWIRE& 104627.83+592843.4 & 1.80 & 1.0 & 45.49 &   8\\  
64 &SWIRE& 104632.93+563530.3 & 1.78 & 1.7 & 45.70 &   8\\  
65 &SWIRE& 104643.29+575851.0 & 1.60 & 1.9 & 45.67 &   8\\  
66 &SWIRE& 104653.08+592652.2 & 1.69 & 2.1 & 45.53 &   8\\  
67 &SWIRE& 104656.24+594008.0 & 1.66 & 1.7 & 45.66 &   8\\  
68 &  SWIRE& 104731.08+581016.1	 & 1.81 & 	1.9 & 45.78 &  7\\
69 &SWIRE& 104754.66+58d3906.0 & 1.63 & 1.6 & 45.61 &  8\\  
70 &  SWIRE& 104839.33+592149.0  & 1.89 & 	2.0 & 45.83 &   7\\
71 &SWIRE& 104843.22+584537.8 & 2.23 & 1.3 & 45.77 &   8\\ 
72 &SWIRE& 104845.20+561055.9 & 1.25 & 2.0 & 45.49 &  8\\ 
73 &SWIRE&104922.65+564032.5 & 1.70 & 1.6 & 45.66 &   8\\ 
74 &SWIRE& 105056.09+562823.0 & 1.54 & 2.1 & 45.68 &   8\\  
75 & IRAS 10485-1447  & 105103.67-150319.8 & 0.133 &\nodata& 44.27 & 13\\
76 & IRAS 10494+4424  &105223.59+440846.5   & 0.092 &\nodata& 44.48 & 13\\ 
77 &SWIRE& 105308.24+591447.5 & 1.47 & 1.8 & 45.58 &   8\\  
78 &SWIRE& 105334.60+574242.3 & 1.65 & 2.8 & 45.88 &   8\\ 
79 &  SWIRE& 105405.49+581400.1 & 	1.82 & 	1.7 & 45.73 &   7\\
80 &SWIRE& 105539.93+571711.9 & 1.70 & 1.6 & 45.66 &   8\\  
81 &SWIRE& 105908.46+574511.4 & 1.65 & 1.6 & 45.63 &   8\\  
82 & IRAS 10565+2448 & 105918.14+243234.3 & 0.043 & \nodata & 44.61&  12\\ 
83 & NGC 3556 &111130.97+554026.8	 & 0.0033 & 320	 & 43.39  & 1\\ 
84 & IRAS 11095-0238  & 111203.34-025424.1  & 0.106 &\nodata& 43.70 & 13\\
85 & IRAS 11130-2659  & 111531.56-271622.7  & 0.136 & \nodata &43.98 & 13\\
86 & NGC 3628 &112017.02+133522.2	 & 0.0024 & 2000	 & 43.41 & 1\\ 
87 &  FSC 11257+5113  &112832.73+505721.1 &0.197 &\nodata &44.23 & 14\\
88 & IRAS 11387+4116  & 114122.04+405950.3   & 0.149 &\nodata & 44.32 & 13\\
89 & IRAS 11506+1331  &115314.17+131426.8   & 0.127 &\nodata & 44.63 & 13\\ 
90 & IRAS 12018+1941 & 120424.53+192509.8	 & 0.169 & \nodata & 44.60&  12\\
91 & NGC 4088 &120534.19+503220.5	 & 0.0032	 & 200	 & 43.13& 1\\ 
92 & IRAS 12071-0444 & 120945.12-050113.9	 & 0.128 & \nodata & 44.55&  12\\
93 & IRAS 12112+0305  &121346.05+020841.3   & 0.073 &\nodata & 44.48 & 13\\ 
94 & NGC 4194&121409.61+543135.9	 & 0.0095	 & 1100	 & 44.03& 1\\
95 & IRAS 12127-1412 & 121519.14-142941.4  & 0.133 &\nodata & 43.54 & 13\\
96 &Mkn 52&122542.67+003420.4	&0.0071	&160	&42.92 & 1\\ 
97 & 3C 273 & 122906.67+020308.1	 & 0.158 & \nodata & 44.86&  12\\ 
98 & SMMJ & 123555.13+620901.6 & 1.88 & 0.5 & 45.23 & 11\\
99 & SMMJ & 123600.16+621047.3 & 2.01 &1.5  &45.76  &   11\\ 
100 & SMMJ & 123616.11+621513.5 & 2.55 & 0.7 & 45.61 &  11\\
101 & SMMJ & 123618.33+621550.4 & 2.00 & 0.6 & 45.36 & 11\\
102 & SMMJ & 123619.13+621004.3 & 2.21 & 0.5 & 45.35 & 11\\
103 & SMMJ & 123621.27+621708.1 & 1.99 &0.65  & 45.39 & 11\\
104 & SMMJ & 123622.66+621629.5 & 1.79 & 0.8 &45.39  &   11\\
105 & SMMJ & 123634.51+621240.9 & 1.23 & 2.5 & 45.58 & 11\\
106 & SMMJ & 123653.22+621116.7 & 0.93  & 0.62 &44.30  & 11\\
107 & SMMJ & 123711.37+621331.1 & 1.98 & 1.0 & 45.57 & 11\\
108 & SMMJ & 123711.97+621325.8 & 1.99 & 0.7 & 45.42 & 11\\
109 & SMMJ & 123707.19+621408.0 & 2.48 & 0.5 & 45.44 &  11\\
110 & IRAS 12359-0725  & 123831.62-074225.8   & 0.138 &\nodata & 44.19 & 13\\ 
111 & NGC 4676 &124610.10+304355.0	 & 0.0222	 & 170	 & 44.02&  1\\ 
112 & IRAS 12514+1027 & 125400.82+101112.4	 & 0.319 & \nodata & 44.92&  12\\
113 & Mkn 231 & 125614.29+565225.1	 & 0.042 & \nodata & 44.27& 12\\
114 & NGC 4818 &125648.90-083131.1	 & 0.0023	 & 720	 & 42.58& 1\\ 
115 & NGC 4945&130527.48-492805.6	 & 0.0009	 & 4500	 & 43.07 & 1\\ 
116 & IRAS 13120-5453 & 131506.40-550923.3	 & 0.031 & \nodata & 44.63&  12\\
117 & IRAS 13335-2612  &133622.29-262734.0   & 0.125 &\nodata & 44.52 & 13\\
118 & IRAS 13342+3932 & 133624.07+391730.1	 & 0.179 & \nodata & 44.95 & 12\\
119 & Mkn 266 &133817.69+481633.9	&0.027	&150	&44.23 &   1\\
120 & Mkn 273 & 134442.12+555313.1	 & 0.038 & \nodata & 44.35&  12\\
121 & IRAS 13509+0442  &135331.57+042804.7    & 0.136 &\nodata & 44.68 & 13\\
122 & IRAS 13539+2920  &135609.99+290535.0    & 0.108 &\nodata & 44.58 & 13\\
123 & IRAS 14060+2919  & 140819.09+290447.2 & 0.117 & \nodata & 44.63 & 13\\
124 & IRAS 14070+0525 & 140931.28+051131.4	 & 0.264 & \nodata & 44.79& 12\\
125  & SST24 & 142500.18+325950.6 & 0.320 &2.5 & 44.37 &  3\\
126 &  SST24&142504.04+345013.7  & 0.0783 & 16.6 &43.96	&   2\\
127 & SST24 & J142538.22+351855.2 & 2.29 & 1.4 & 45.83 & 9\\ 
128 &  SST24&142543.59+334527.6  & 0.0715 & 12.3 &43.75	&   2\\ 
129 &  SST24&142554.57+344603.2  &0.0350 & 38.6 &43.62	&   2\\
130   & SST24 &  142607.93+345045.4 & 0.408 & 5.0 & 44.90 &  3\\
131    & SST24 & 142618.65+345703.9 & 0.42 & 0.64 & 44.04&  3\\
132 & SST24 & 142626.49+344731.2 & 2.09 & 2.1 & 45.94 & 9\\ 
133 &  SST24&142629.15+322906.7 & 0.1028 & 6.9 & 43.82 &  2\\ 
134 &  SST24&142646.63+322125.6 & 0.2683 & 9.2&44.81 &  2\\ 
135 &  SST24 & 142651.9+343135	 & 0.51	 & 1.4	 & 44.53 & 6\\
136 &  SST24&142658.80+333314.0  & 0.1548 & 16.4 & 44.56	& 2\\ 
137 &  SST24&142659.15+333305.1  & 0.1514 & 13.0 & 44.44	&   2\\ 	
138  &  SST24& 142732.9+324542	 & 0.37	 & 1.2	 & 44.18 &   6\\
139 & IRAS 14252-1550  & 142801.02-160339.5  & 0.149 &\nodata & 44.25 & 13\\
140   & SST24 & 142819.51+335150.1 & 0.493 & 4.0 & 44.98&  3\\
141  & SST24 &  142849.79+343240.3 & 0.219 & 16.3 & 44.84 &  3\\
142    & SST24 &  142919.77+341506.2 & 0.424 & 2.6 & 44.65 &  3\\
143   & SST24 &  142935.97+333713.4 & 0.424 & 3.3 & 44.76 &  3\\
144  & SST24 &  143000.39+353814.6 & 0.243 & 7.1 & 44.58&  3\\
145  &SST24 & 143024.46+325616.4 &0.040  &22.2  &43.48  & 3\\ 
146 &  SST24& 143039.27+352351.0 &0.0885 & 14.4 &44.01	&  2\\
147  & SST24 &  143114.77+334623.0 & 0.230 & 7.6 & 44.56	 &  3\\
148 &  SST24& 143115.23+324606.2 & 0.0244 & 9.8 &42.71&  2\\
149 &  SST24&143119.79+353418.1  &0.0333& 39.0 &43.58	& 2\\ 
150 &  SST24& 143120.00+343804.2 &0.0156 & 2.7 &41.76	&  2\\
151 &  SST24&143121.15+353722.0  & 0.0347 & 51.9  &43.74	&   2\\
152 &  SST24&143125.46+331349.8  & 0.0233 & 15.1 &42.85 &   2\\
153 &  SST24& 143126.81+344517.9 & 0.0835 & 13.7 &43.94	&  2\\
154  &  SST24  & 143151.8+324327	 & 0.66	 & 3.6	 & 45.19 &  6\\ 
155 &  SST24& 143156.25+333833.3  & 0.0329 & 27.8 &43.43	&   2\\
156  &   SST24 & 143218.1+341300	 & 0.98	 & 2.2	 & 45.32 &   6\\ 
157 &SST24 & 143228.36+345838.8 &0.1312  & 5.2 &43.90  & 3\\ 
158 & SST24 & 143234.90+332832.3 &0.2506  &4.2  &44.39  & 3\\ 
159 & SST24 & 143326.18+330558.6 &0.2426  &4.1  &44.35  & 3\\  
160  & SST24 & 143445.32+331346.1 &0.0762  &7.7  &43.46  & 3\\  
161  &  SST24  & 143449.3+341014	 & 0.50	 & 2.4	 & 44.74 &  6\\ 
162    & SST24 & 143453.85+342744.4 & 0.34 & 0.97 & 44.02   & 3\\
163   & SST24 &  143606.84+350927.8 & 0.54 & 1.4 & 44.60 &  3\\
164 & SST24 & 143619.14+332917.9 &0.1886  &4.9  &43.37  &  3\\  
165 & SST24 & 143628.12+333358.0 &0.2642  &7.8  &44.66  & 3\\ 
166 &  SST24& 143632.01+335230.7 &0.0883 & 9.4 &43.82	&   2\\
167 & SST24 & 143636.65+345034.0 &0.2781  &2.4 &44.19  & 3\\ 
168  &  SST24  & 143639.0+345222	 & 0.98	 & 2.4	 & 45.37 &  6\\ 
169 &  SST24& 143641.26+345824.4 & 0.0290 & 21.8 &43.21	&  2\\

170 & IRAS 14348-1447 & 143738.27-150024.6	 & 0.083 & \nodata & 44.47 &  13\\
171  &  SST24  & 143820.7+340233	 & 0.66	 & 6.0	 & 45.40 &  6\\
172 & IRAS 14378-3651 & 144058.90-370433.0	 & 0.068 & \nodata & 44.38&  12\\
173 &  FSC 14475+1418 &144954.86+140610.5&0.251 & \nodata &44.53& 14\\
174 &  FSC 14516+3851 &145335.96+383913.1 &0.153 &\nodata & 43.85& 14\\
175 & IRAS 15001+1433 & 150231.94+142135.3	 & 0.163 & \nodata & 44.91&  12\\
176 & IRAS 15206+3342  & 152238.12+333136.1   & 0.125 &\nodata & 44.78 & 13\\
177 & IRAS 15225+2350  & 152443.94+234010.2 & 0.139 &\nodata & 44.40 & 13\\
178 & IRAS 15250+3609 & 152659.40+355837.5	 & 0.055 & \nodata & 44.09& 12\\
179 & Arp 220  & 153457.24+233011.7  & 0.018 &\nodata & 43.87 & 13\\
180 &  FSC 15385+4320 & 154014.02+431042.4 &0.380 &\nodata &44.48 & 14\\
181 & IRAS 15462-0450 & 154856.80-045933.7	 & 0.100 & \nodata & 44.36& 12\\
182 &  FSC 15585+4518 &160003.29+451046.2 &0.486  &\nodata & 44.95& 14\\
183 & FSC 16073+0209 &160949.75+020130.8 & 0.223 &\nodata &44.72 & 14\\
184 & IRAS16090-0139 & 161140.42-014705.8	 & 0.134 & \nodata & 44.61 & 13\\
185 &SWIRE& 161744.64+540031.4 & 1.80 & 1.5 & 45.66 &    8\\
186 & IRAS 16468+5200  &164801.70+515544.4  & 0.150 &\nodata & 44.16 & 13\\ 
187 & IRAS 16474+3430  & 164914.20+342510.1  & 0.111 &\nodata &44.66 & 13\\
188 & IRAS 16487+5447  &164946.88+544235.4    & 0.104 &\nodata & 44.14 & 13\\
189 & NGC 6240 & 165258.89+022403.4	 & 0.024 & \nodata & 44.00&  12\\
190 & IRAS 17028+5817  & 170341.91+581344.4   & 0.106 &\nodata & 44.36 & 13\\ 
191 & IRAS 17044+6720  &170428.41+671628.5  & 0.135 & \nodata &44.29 & 13\\ 
192 & MIPS 506 &  171138.59+583836.7 & 	2.52 & 	2.2 & 46.10 &  10\\ 
193 & MIPS 8184 &  171226.76+595953.5 & 	0.99 & 	1.4 & 45.14 & 10\\ 
194  & SST24 & 171239.64+584148.3 & 0.166 &11.5 & 44.45 &  5\\
195    & SST24 & 171308.57+601621.0 & 0.334 & 5.2 & 44.74&  5\\
196 & MIPS 289 &  171350.00+585656.8 & 	1.86 & 	2.1 & 45.84 &   10\\ 
197   & SST24 &  171427.02+583836.7 & 0.57 & 1.7 & 44.73&  5\\
198  & SST24 & 171447.76+593509.6 & 0.204 & 6.1 & 44.36 &   5\\
199 & MIPS 283 &  171458.27+592411.2 & 	0.94 & 	1.4 & 45.06 & 10\\ 
200  & SST24 &  171525.74+600424.1 & 0.50 & 2.3 & 44.75&  5\\ 
201  & SST24 &  171542.00+591657.4 & 0.119 & 12.0 & 44.17 &   5\\
202    & SST24 &  171555.23+601348.0 & 0.43 & 2.5 & 44.65 &   5\\
203 & SST24 & 171607.21+591456.3 & 0.0544 & 8.8 &43.06 & 4\\
204& MIPS 429 &  171611.81+591213.3 & 	2.09 & 	1.4 & 45.76 & 10\\ 
205 & SST24 & 171634.03+601443.6 & 0.1057 & 3.9 & 43.29 & 4\\
206 & SST24 & 171641.09+591857.2 & 0.0570 & 6.1 & 42.94 & 4\\
207 & MIPS 8493 &  171805.10+600832.7 & 	1.80 & 	1.4 & 45.64 &  10\\ 
208 & MIPS 22277 &  171826.67+584242.1 & 	1.77 & 	2.8 & 45.93 &  10\\ 
209 & IRAS 17179+5444 & 171854.23+544147.3	 & 0.147 & \nodata & 44.38&  12\\
210 & MIPS 15928 &  171917.45+601519.9 & 	1.52 & 	1.5 & 45.53 &  10\\
211 & MIPS 22651 &  171926.47+590929.9 & 	1.73 & 	1.5 & 45.64 & 10\\ 
212 & MIPS 16030 &  172000.32+601520.9 & 	0.98 & 	0.7 & 44.83 & 10\\ 
213   & SST24 & 172025.19+591503.5 & 0.303 & 4.7 & 44.60 &  5\\
214 & MIPS 22554 &  172059.80+591125.7 & 	0.82 & 	1.3 & 44.94 & 10\\ 
215 & MIPS 22482 &  172100.39+585931.0 & 	1.84 & 	1.6 & 45.71 & 10\\ 
216 & MIPS 22600 & 172218.34+584144.6 & 	0.86 & 	0.9 & 44.83 & 10\\ 
217 & SST24 & 172220.27+590949.7 & 0.180 &6.4 &  44.27 &   5\\
218 & MIPS 22530 &  172303.30+591600.2 & 	1.96 & 	1.8 & 45.82 &  10\\ 
219 & IRAS 17208-0014 & 172321.93-001700.4	 & 0.043 & \nodata & 44.62 & 12\\
220   & SST24 & 172348.13+590154.5 & 0.318 & 3.6 & 44.53&  5\\
221  & SST24 &  172400.61+590228.2 & 0.1787 & 6.2 & 44.25 &   5\\
222 & MIPS 16144 &  172422.10+593150.8 & 	2.13 & 	1.8 & 45.88 & 10\\ 
223   & SST24 &  172503.37+591109.1 & 0.513 & 2.0 & 44.71 &   5\\
224 & SST24 & 172540.26+584953.4 &	1.77 &	2.2 & 45.83 &  9\\ 
225 & SST24 & 172611.96+592851.8 & 0.0729 & 6.4 & 43.18 & 4\\
226 & IRAS 19254-7245 & 193122.51-723920.2	 & 0.063 & \nodata & 44.47&  12\\
227 & IRAS 19297-0406 & 193221.25-035956.3	 & 0.086 & \nodata & 44.70&  12\\
228 & IRAS 20087-0308 & 201123.86-025950.8	 & 0.106 & \nodata & 44.83&  12\\
229 & IRAS 20100-4156 & 201329.85-414734.7	 & 0.130 & \nodata & 44.81& 12\\
230 & IRAS 20414-1651 & 204418.18-164016.4  & 0.086 &\nodata & 44.20 & 13\\
231 & IRAS 20551-4250 & 205826.78-423901.6	 & 0.043 & \nodata & 44.34&  12\\
232 & IRAS 21208-0519  &212329.19-050656.4     & 0.130 &\nodata & 44.48 & 13\\
233 & IRAS 21329-2346  &213545.88-233234.7  & 0.125 &\nodata & 44.17 & 13\\
234 & NGC 7252 &222044.77-244041.8	 & 0.0157	 & 350	 & 43.92&  1\\
235 & IRAS 22206-2715  & 222328.89-270003.1   & 0.132 &\nodata & 44.19 & 13\\ 
236 & IRAS 22491-1808  & 225149.35-175224.0  & 0.076 & \nodata &44.29 & 13\\
237 & IRAS 23128-5919 & 231547.01-590316.9	 & 0.045 & \nodata & 44.44& 12\\
238 & IRAS 23234+0946  & 232556.15+100250.9    & 0.128 &\nodata & 44.19 & 13\\
239 & IRAS 23230-6926 & 232603.57-691020.3	 & 0.106 & \nodata & 44.62&  12\\
240 & IRAS 23253-5415 & 232806.10-535831.0	 & 0.130 & \nodata & 44.51& 12\\
241 & IRAS 23327+2913  & 233511.93+293000.2   & 0.107 & \nodata &44.09 & 13\\
242 & NGC 7714 &233614.10+020918.6& 0.0090	 & 400	 & 43.56& 1\\ 
243 & IRAS 23498+2423 & 235226.05+244016.2	 & 0.212 & \nodata & 44.70 & 12\\
\enddata
\tablenotetext{a}{Observed flux density in mJy at peak of 7.7$\mu$m PAH feature measured from published spectra or from our own extractions. Typical uncertainties are $\pm$ 5\% for sources with f$_{\nu}$(7.7$\mu$m) $\ga$ 5 mJy and $\pm$ 10\% for sources with f$_{\nu}$(7.7$\mu$m) $\sim$ 1 mJy. Sources without entries have $\nu$L$_{\nu}$(7.7$\mu$m) determined from other PAH features, as explained in footnote b.}
\tablenotetext{b}{Rest frame luminosity $\nu$L$_{\nu}$(7.7$\mu$m) in ergs s$^{-1}$ determined using peak f$_{\nu}$(7.7$\mu$m) and luminosity distances from E.L. Wright, http://www.astro.ucla.edu/~wright/CosmoCalc.html, for H$_0$ = 71 \kmsMpc, $\Omega_{M}$=0.27 and $\Omega_{\Lambda}$=0.73. For ULIRGs in reference 12, $\nu$L$_{\nu}$(7.7\,\um) is determined using the relation $\nu$L$_{\nu}$(7.7$\mu$m) = 28 L(6.2$\mu$m+11.3$\mu$m); for ULIRGs in references 13 and 14, $\nu$L$_{\nu}$(7.7$\mu$m) is determined using the relation $\nu$L$_{\nu}$(7.7$\mu$m) = 50 L(6.2$\mu$m). For starbursts in reference 1, luminosities also include the correction factor FF given in \citet{bra06} for flux from the resolved source which is not included in the IRS observing slit.}
\tablenotetext{c}{1 = \citet{bra06}; 2 = \citet{hou07}; 3 = new extraction of Bootes source from archival program 20128 (G. Lagache, P.I.) or 20113 (H. Dole, P.I.); 4 = new extraction from program 40038 (Weedman and Houck, in preparation);  5 = new extraction of FLS source from archival program 20128;  6 = \citet{brn08}; 7 = \citet{wee06a}; 8 = \citet{far08}; 9 = new extraction of source in \citet{hou05} or \citet{wee06b} using v15 data products and the 4 pixel extraction technique described in \citet{wee06a}; 10 = \citet{yan07}; 11 = \citet{pop08}; 12 = \citet{far07}; 13 = \citet{ima07}; 14 = \citet{sar08}.}
\end{deluxetable}

\clearpage
%
%
\begin{figure}
\figurenum{1}
\includegraphics[scale=1.0]{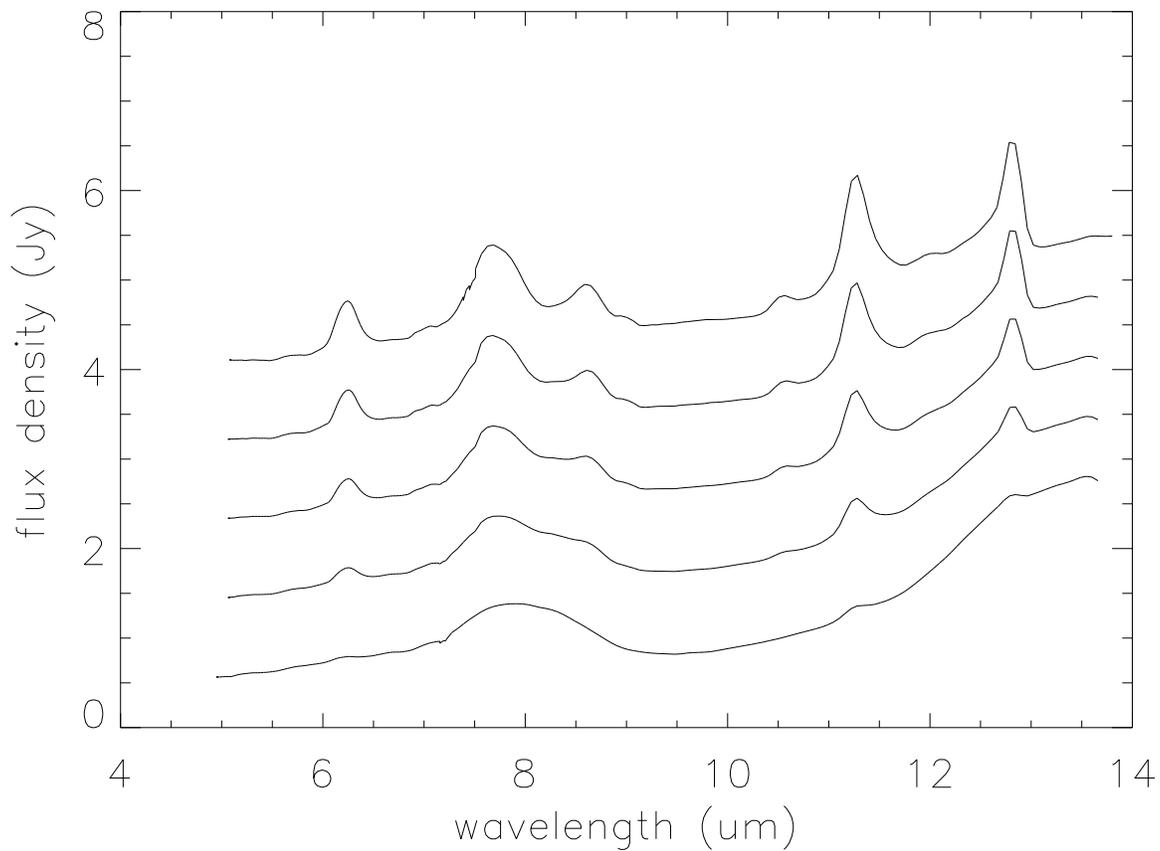}
\caption{Rest-frame spectra of starburst NGC 7714 (top) and AGN Markarian 231 (bottom); intermediate spectra are synthetic composite spectra combining starburst and AGN with 75\%, 50\%, and 25\% starburst components (top to bottom).  All spectra are normalized to the peak which falls between 7.7$\mu$m and 7.9$\mu$m, depending on the composite.  The spectrum of Markarian 231 is in correct flux units with correct zero point.  Other spectra are normalized and have zero points successively displaced by 1 Jy so that spectra do not overlap.} 

\end{figure}

\begin{figure}
\figurenum{2}
\includegraphics[scale=0.90]{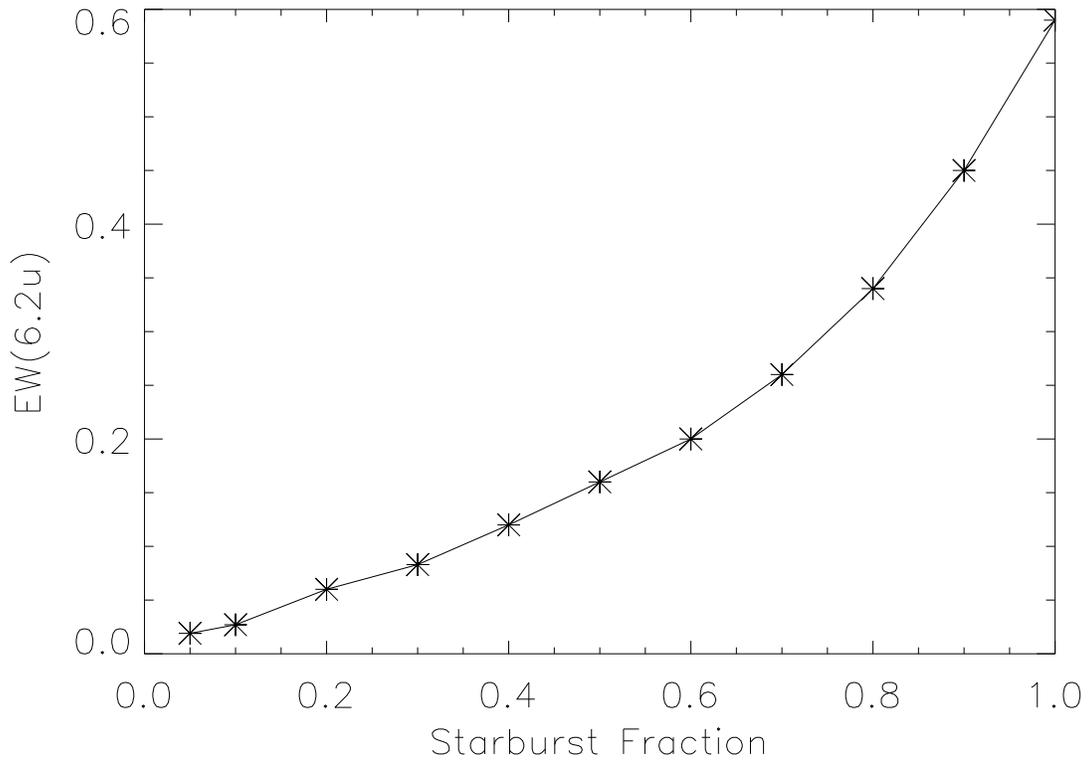}
\caption{Equivalent width of 6.2\,\um PAH feature in source rest frame compared to fraction of composite spectrum that arises from starburst like NGC 7714, using AGN and starburst composite spectra as in Figure 1.}

\end{figure}


\begin{figure}
\figurenum{3}
\includegraphics[scale=0.90]{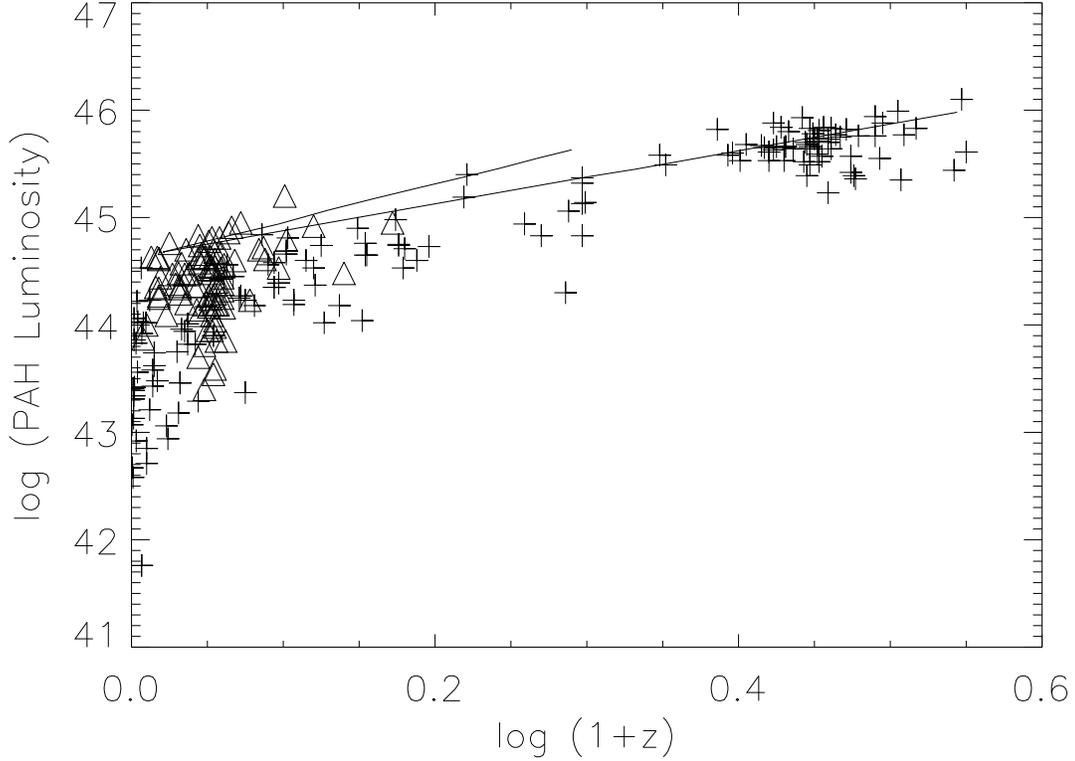}
\caption{Starburst PAH luminosity $\nu$L$_{\nu}$(7.7$\mu$m) in ergs s$^{-1}$ compared to redshift for all sources in Table 1. Crosses are luminosities from measured $\nu$L$_{\nu}$(7.7$\mu$m) for pure starbursts; triangles are values for starburst component of IRAS ULIRGS determined from luminosities in 6.2\,\um and 11.3\,\um PAH features and changed to $\nu$L$_{\nu}$(7.7$\mu$m) by $\nu$L$_{\nu}$(7.7$\mu$m) = 28L(6.2$\mu$m+11.3$\mu$m) for reference 12 or $\nu$L$_{\nu}$(7.7$\mu$m) = 50L(6.2$\mu$m) for references 13 and 14. Upper solid line is luminosity evolution to z = 1 by factor (1+z)$^{3.5}$ from \citet{lef05}; lower solid line is luminosity evolution to z = 2.5 by factor (1+z)$^{2.5}$, determined from best fit to the most luminous sources shown.  With assumptions discussed in text, PAH luminosity shown scales to bolometric luminosity $L_{ir}$ by log $L_{ir}$ = log[$\nu$L$_{\nu}$ (7.7$\mu$m)] + 0.78, and scales to star formation rate by log[SFR] = log[$\nu$L$_{\nu}$ (7.7$\mu$m)] - 42.57, for SFR in units of \mdot. Luminosity uncertainties arising from flux uncertainties are $\pm$ 5\% to $\pm$ 10\%, depending on source flux, which are smaller than plotted symbols.} 

\end{figure}

\begin{figure}
\figurenum{4}
\includegraphics[scale=0.90]{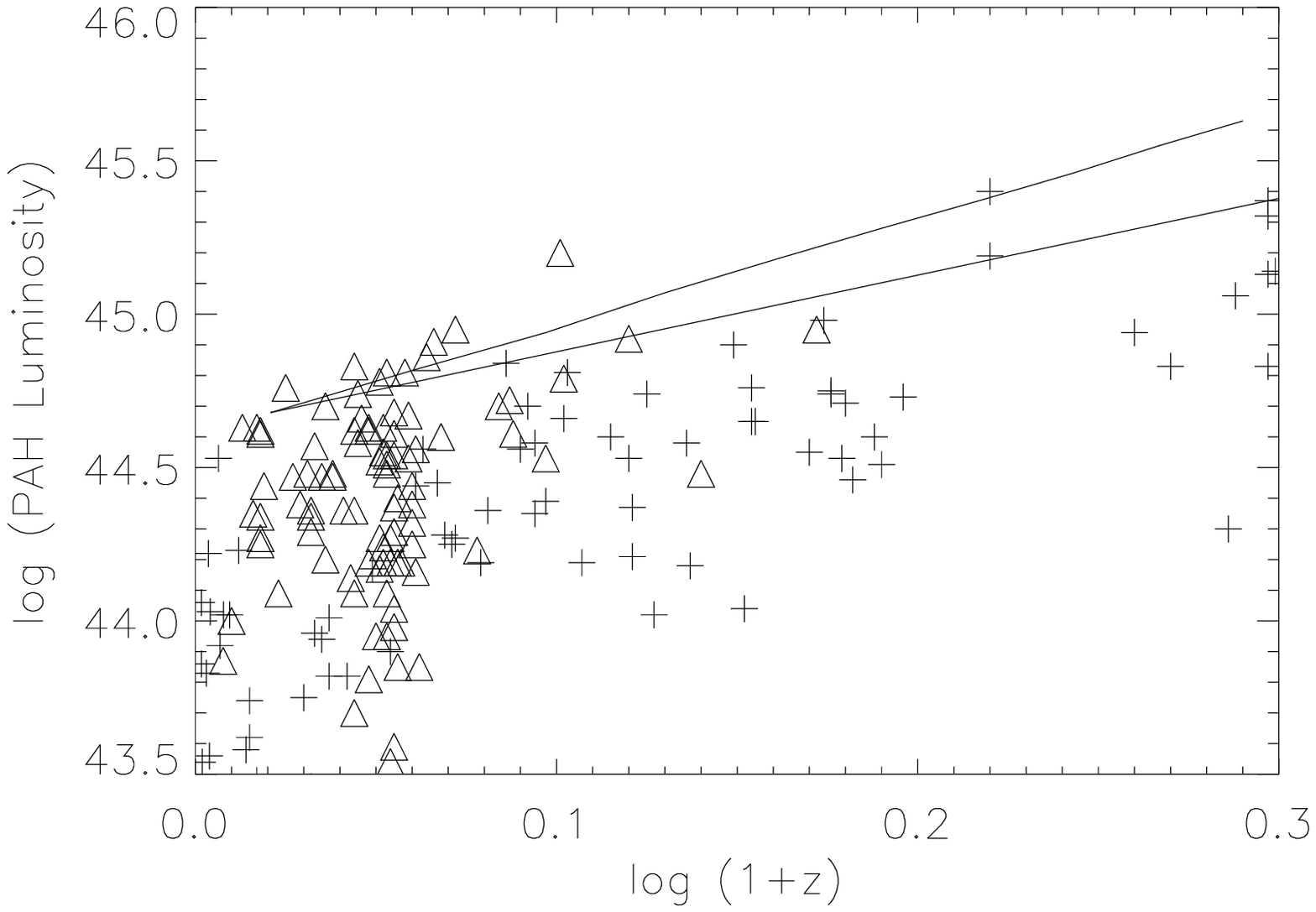}
\caption{Starburst PAH luminosity $\nu$L$_{\nu}$(7.7$\mu$m) in ergs s$^{-1}$ compared to redshift for all sources in Table 1 with z $<$ 1.0 and log[$\nu$L$_{\nu}$(7.7$\mu$m)] $>$ 43.5. Crosses are luminosities from measured $\nu$L$_{\nu}$(7.7$\mu$m) for pure starbursts; triangles are values for starburst component of IRAS ULIRGS determined from luminosities in 6.2\,\um and 11.3\,\um PAH features and changed to $\nu$L$_{\nu}$(7.7$\mu$m) by $\nu$L$_{\nu}$(7.7$\mu$m) = 28L(6.2$\mu$m+11.3$\mu$m) for reference 12 or $\nu$L$_{\nu}$(7.7$\mu$m) = 50L(6.2$\mu$m) for references 13 and 14. Upper solid line is luminosity evolution to z = 1 by factor (1+z)$^{3.5}$ from \citet{lef05}; lower solid line is luminosity evolution to z = 2.5 by factor (1+z)$^{2.5}$, determined from best fit to the most luminous sources over all redshifts in Figure 3.  With assumptions discussed in text, PAH luminosity shown scales to bolometric luminosity $L_{ir}$ by log $L_{ir}$ = log[$\nu$L$_{\nu}$ (7.7$\mu$m)] + 0.78, and scales to star formation rate by log[SFR] = log[$\nu$L$_{\nu}$ (7.7$\mu$m)] - 42.57, for SFR in units of \mdot.  Luminosity uncertainties arising from flux uncertainties are $\pm$ 5\% to $\pm$ 10\%, depending on source flux, which are smaller than plotted symbols.} 

\end{figure}


\begin{thebibliography}

\bibitem [Armus et al. (2007)]{arm07}
Armus, L. et al. 2007, \apj, 656, 148
\bibitem [Bavouzet et al. (2007)]{bav07}
Bavouzet, N., Dole, H., Le Floc'h, E., Caputi, K.I., Lagache, G., and Kochanek, C.S. 2007, \aap, in press (astro-ph/07120965)
\bibitem [Brand et al. (2008)]{brn08}
Brand, K., et al. 2008, \apj, in press (astro-ph/07093119)
\bibitem [Brandl et al.(2006)]{bra06}
Brandl, B. et al. 2006, \apj, 653, 1129.
\bibitem [Calzetti et al. (2007)]{cal07}
Calzetti, D. et al. 2007, \apj, in press (astro-ph/07053377)
\bibitem [Caputi et al. (2007)]{cap07}
Caputi, K.I., et al. 2007, \apj, in press (astro-ph/0701283)
\bibitem [Chapman et al. (2005)]{cha05}
Chapman, S. C., Blain, A. W., Smail, Ian, and Ivison, R. J. 2005, \apj, 622, 772
\bibitem [Desai et al. (2007)]{des07}
Desai, V., et al. 2007, \apj, in press (astro-ph/07074190)
\bibitem [Draine and Li (2001)]{dra01}
Draine, B. T. and Li, A. 2001, \apj, 551, 807
\bibitem [Fadda et al. (2006)] {fad06}
Fadda, D. et al. 2006, \aj, 131, 2859 
\bibitem [Farrah et al. (2007)]{far07}
Farrah, D., et al. 2007, \apj, in press (astro-ph/07060513)
\bibitem [Farrah et al. (2008)]{far08}
Farrah, D., et al. 2007, \apj, in press (astro-ph/08011842)
\bibitem [Fazio et al. (2004)]{faz04}
Fazio, G. et al. 2004, \apjs, 154, 10
\bibitem [F\"{o}rster Schreiber et al.(2004)]{for04}
F\"{o}rster Schreiber, N. M., Roussel, H., Sauvage, M., and Charmandaris, V. 2004, \aap, 419, 501
\bibitem [Genzel et al.(1998)]{gen98}	
Genzel, R. et al. 1998, \apj, 498, 579
\bibitem [Haarsma et al. (2000)] {haa00}
Haarsma, D. B., Partridge, R. B., Windhorst, R. A.,and Richards, E. A. 2000, \apj, 544, 641
\bibitem [Higdon et al.(2004)]{hig04}
Higdon, S.J.U., et al. 2004, \pasp, 116, 975
\bibitem [Ho and Keto (2007)]{ho07}
Ho, L, and Keto, E. 2007, \apj, in press (astro-ph/0611856)
\bibitem[Hopkins (2004)]{hop04}
Hopkins, A.M. 2004, \apj, 615, 209
\bibitem [Houck et al.(2004)]{hou04} 
Houck, J. R., et al. 2004, \apjs, 154, 18
\bibitem [Houck et al.(2005)]{hou05}
Houck, J.R., et al. 2005, \apjl, 622, L105 (H05)
\bibitem [Houck et al. (2007)]{hou07}
Houck, J.R., Weedman, D.W., Le Floc'h, E., and Hao, L. (2007), \apj, in press (astro-ph/07082400)
\bibitem [Imanishi et al. (2007)]{ima07}
Imanishi, M., Dudley, C. C., Maiolino, R., Maloney, P. R., Nakagawa, T., and Risaliti, G. 2007, \apj, in press, (astro-ph/0702136)
\bibitem [Kennicutt (1998)]{ken98}
Kennicutt, R.C. 1998, Ann.Rev.Astron.Ap, 36, 189
\bibitem [Lagache et al. (2004)]{lag04}
Lagache, G. et al. 2004, \apjs, 154, 112
\bibitem [Le Floc'h et al.(2005)]{lef05}
Le Floc'h, E. et al. 2005, \apj, 632, 169
\bibitem [Levenson et al. (2007)]{lev07}
Levenson, N.A., Sirocky, M.M., Hao, L., Spoon, H.W.W., Marshall, J.A., Elitzur, M., and Houck, J.R. 2007, \apjl, 654, L45
\bibitem [Lonsdale et al. (2004)]{lon04}
Lonsdale, C. et al. 2004, \apjs, 154, 54
\bibitem [Madau et al.(1998)]{mad98}
Madau, P., Pozzetti, L., and Dickinson, M. 1998, \apj, 498, 106
\bibitem [Mannucci et al.(2007)]{man07}
Mannucci, F., Buttery, H., Maiolino, R., Marconi, A., and Pozzetti, L. 2007, \aap, 461, 423
\bibitem [Martin et al. (2008)]{mar08}
Martin, D.C. et al. 2008, \apjs, in press (astro-ph 07090730)
\bibitem [Peeters et al. (2004)]{pee04}
Peeters, E., Spoon, H.W.W., and Tielens, A.G.G.M. 2004, \apj, 613, 986
\bibitem[Pope et al.(2008)]{pop08}
Pope, A. et al., 2008, \apj, in press (astro-ph/07111553) 
\bibitem [Rieke et al.(2004)]{rie04}
Rieke, G.H. et al., 2004, \apjs, 154, 25
\bibitem [Rigopoulou et al.(2000)]{rig00}	
Rigopoulou, D., Spoon, H. W. W., Genzel, R., Lutz, D., Moorwood, A. F. M., and Tran, Q. D. 2000, \aj, 118, 2625	
\bibitem [Sargsyan et al. (2008)]{sar08}
Sargsyan, L., Mickaelian, A., Weedman, D., and Houck, J. 2008, \apj, submitted
\bibitem [Spoon et al.(2007)]{spo07}
Spoon, H.W.W. et al 2007, \apjl, 654, L49
\bibitem [Takeuchi et al.(2005)]{tak05}
Takeuchi, T. T., Buat, V., Iglesias-Paramo, J., Boselli, A., and Burgarella, D. 2005, \aap, 432, 423
\bibitem [Weedman et al.(2005)]{wee05}
Weedman, D.W., et al. 2005, \apj, 633, 706
\bibitem [Weedman et al.(2006a)]{wee06a}
Weedman, D.W., et al., 2006, ApJ, 653, 101.
\bibitem [Weedman et al.(2006b)]{wee06b}
Weedman, D.W., Le Floc'h, E., Higdon, S.J.U., Higdon, J.L., and Houck, J.R. 2006, \apj, 638, 613
\bibitem [Yan et al.(2007)]{yan07}
Yan, L. et al. 2007, \apj, 658, 778

\end{thebibliography}
\end{document}